\documentclass[11pts,twocolumn]{article}
\usepackage[dvips]{graphics}
\usepackage{epsfig}
\usepackage[centertags]{amsmath}
\usepackage{amsfonts}
\usepackage{amssymb}
\usepackage{amsthm}
\usepackage{newlfont}

\begin{document}
\title{Lawn tennis balls, Rolling friction experiment and Trouton viscosity }
\author{Ankit Singh, Devang Jain, Kaustubh Kulkarni, Abhishek Budhraja, K. R. Sai
Tej,\\
Satish Sankarlingam ;
Anindya Kumar Biswas, Physics;\\
BITS-Pilani Goa Campus, Vasco, Goa, Pin-403726.\\
email:anindya@bits-goa.ac.in}
\date{\today}
\maketitle
 \begin{abstract}
We took three lawn tennis balls arbitrarily. One was moderately
old, one was old and another was new. Fabricating a conveyor belt
set-up we have measured rolling friction coefficients, $\mu_{r}$,
of the three balls as a function of their angular velocities,
$\omega_{ball}$. In all the three cases, plotting the results and
using linear fits, we have obtained relations of the form
$\mu_{r}= k_{rol} \omega_{ball}$ and have deduced the
proportionality constant $k_{rol}$. Moreover, core of a lawn
tennis ball is made of vulcanised India-rubber. Using the known
values of Young modulus and shear viscosity of vulcanised
India-rubbers in the theoretical formula for $k_{rol}$, we
estimate $k_{rol}$s for the cores made of vulcanised
India-rubbers, assuming Trouton ratio as three. The experimental
results for the balls and the semi-theoretical estimates for the
cores, of $k_{rol}$, are of the same order of magnitudes.
\end{abstract}

\begin{section}{Introduction}
The game of tennis or, lawn-tennis as it's known nowadays, has an
interesting history. Even the balls \cite{{itf},{tennisdesign}}
used in the game also enjoys an equally interesting story. A ball
consists of two parts. The core is made of vulcanised
India-rubber\cite{vulcan}, whereas, the cover is made of seamless
fabric, composed of wool and artificial fiber. The ball has
evolved through time and today all balls come in yellow colour.

We randomly took three lawn tennis balls. One was moderately old,
after having been played. Another was old, wool was almost not
there. The third ball was brand new. Primary objective was to do a
rolling friction experiment. Main motivation behind, was to do
phenomenology, namely check the formula of Brilliantov
et.al.\cite{brilliant}, relating rolling friction coefficient with
angular velocity.  We embark on measuring rolling friction
coefficient of these three balls with the instrument we made,
following Ko. et.al.\cite{ko}. Ko. et.al. developed their working
principle of measurement, assuming the presence of rolling
friction. They did it keeping the linear result of Brilliantov
et.al.\cite{brilliant} on bulk rolling friction, in mind.

Friction is something that always resists the motion of any
object. Friction is of mainly three types-static, sliding and
rolling. While static and sliding frictions are forces, mainly due
to surface effects and act through the point of contact of two
bodies, rolling friction is predominantly a bulk
effect\cite{brilliant}, with very little effect, arising from the
surface of contact. Unlike the other two, rolling friction is a
couple. It opposes rotational motion only. It is due to hysteresis
loss during repeated cycles of deformation and recovery of the
body undergoing rolling. Rolling friction mainly depends upon four
parameters. These are mechanical properties of the body, radius of
the rolling object, nature of the ground and the forward speed
respectively. Mechanical properties involved are elastic constants
and viscosity coefficients.

The measurement of rolling friction, in general, follows the
following recent story. In a paper by Domenech et.al.\cite{dom},
it was found that the coefficient of rolling friction of a ball
rolling up an incline is dependent on the radius of the ball. To
measure the speed dependent coefficient of rolling friction,
Edmonds et.al.\cite{Ed} constructed a set up analogous to a
cyclotron used to accelerate protons. However, the measurements
require the availability of relevant data such as frictions at
different joints and the relation between the coefficient of
rolling friction and the speed of the ball. Soodak and
Tiersten\cite{Soo} used the perturbation method to analyze effects
of rolling friction on the path of a rolling ball on a spinning
turntable. Similar effects were also analysed by Ehrlich and
Tuszynski\cite{ehr} who proposed to measure the coefficient of
rolling friction of the ball when it starts to rotate, using an
inclined mirror. The coefficient of rolling friction was found to
be tangent of the angle of inclination. Budinski\cite{bud} used
the similar method to measure the break away coefficient of
rolling friction for rolling element bearings. \\
The set-up proposed by Y. Xu, K. L. Yung and S. M. Ko\cite{ko}, in
particular, consists of a wood plate and a plastic frame. The
plastic frame was mounted around the belt to keep the ball in
place. The conveyor belt and the motor were fixed on the wood
plate of which one end was supported by the jack. The angle of
inclination of the belt was varied by changing the height of the
jack. Rotational speed of the squash ball used, was between twenty
to forty rpm. They got the value of $k_{rol}$ as $0.0005$ minutes
while the full relation obtained experimentally was
$\mu_{r}(\omega)=0.0005\omega+0.0092$ with linear regression
coefficient $0.96734$. \\This paper is organised as follows. In
the section II, we describe our setup and the working principle
behind our measurement. In the section III, we develop the
theoretical background required to interpret our experimental
results. In the next section IV, we describe our experiment and
our observations with analysis. In the section V. we compare our
experimental results with that deduced semi-theoretically,
followed by a discussion on different issues and possibilities
arising out of the experiment.
\end{section}
\begin{section}{Experimental Setup and Working Principle}
\begin{figure}
%\centering
\includegraphics{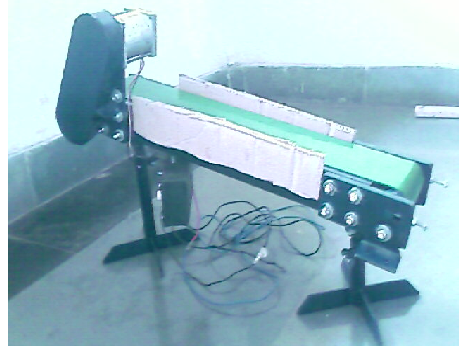}
\caption{Conveyor belt set-up}
\includegraphics{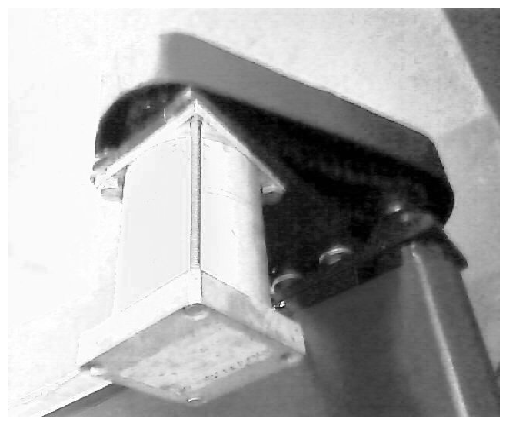}
\caption{DC motor of the set-up}\label{Figure2}
\end{figure}
The conveyor belt setup was made of steel to provide stability at
higher rpms of the motor. The motor was connected to the rollers
of the belt using a gear chain arrangement. The motor was also
provided with an electric dimmer to vary the rotational speed of
the motor. The conveyor belt was mounted on stands of variable
heights on either end to facilitate variation in the angle of the
inclined surface of the belt. The belt was 12.7 cm wide and 140 cm
long. The diameters of the rollers of the conveyor belt were 4.36
cm. In place of the hollow sphere, tennis balls of varied texture,
age and weight were used. Temperature conditions of $32^{0}$C,
high humidity and negligible wind conditions were maintained
throughout the experiment.\\
The conveyor belt assembly consists of a dc motor and an
electrical setup for converting ac supply voltage to the required
dc voltage. The dc motor is of 0.25 hp and peak speed of around
150 rpm.

\noindent Fig.\ref{Figure3} depicts a viscoelastic sphere of
radius $r$ rolling down an incline of angle $\beta$.
\begin{figure}
%\centering
\includegraphics{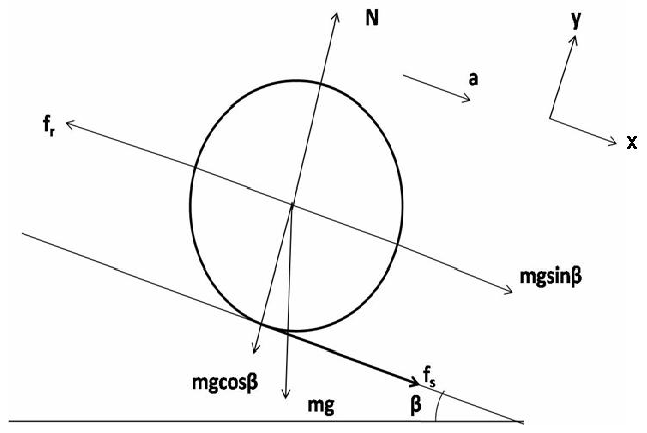}
\caption{a viscoelastic sphere rolling down a hard
incline}\label{Figure3}
\end{figure}
\noindent Here, $mg$ is the weight of the body, $f_{s}$ is the
sliding friction force plus $f_{r}$, where, $f_{r}$ is the rolling
friction force, $a$ is the acceleration of the sphere
\underline{with respect to the ground}, along the incline. We
recall that the rolling friction is a couple. Assuming the
co-ordinate axis as shown in the Fig.\ref{Figure3}, and applying
Newton's second law along the incline, we get
\begin{equation}\label{e:tr}
ma = mgsin\beta  + f_{s} - f_{r}
\end{equation}
and normal to the axis we get
\begin{equation}\label{e:n}
 N = mg cos\beta
\end{equation}
Taking torque about the center of the sphere, we get
\begin{equation}\label{e:fs}
\tau = rf_{s}
\end{equation}
where $\tau$ is the net torque. But we know that torque of a body
can be written as
\begin{equation}\label{e:al}
\tau = I\alpha
\end{equation}
where $I$ is the moment of inertia of the body and $\alpha$ is the
angular acceleration of the body w.r.t its center of mass.
Equating the two equations(\ref{e:fs},\ref{e:al}), we get
\begin{equation}
I\alpha =  rf_{s}
\end{equation}
or,
\begin{equation}\label{e:tor}
f_{s} = \frac{I\alpha}{r}
\end{equation}
For rolling,
\begin{equation}
 v = r \omega_{ball}
\end{equation}
Equivalently this is
\begin{equation}
a=r\alpha
\end{equation}
Substituting this in the eq.(\ref{e:tor}), we get
\begin{equation}\label{e:fr}
 f_s = \frac{Ia}{r^{2}}
\end{equation}
Combining eq.(\ref{e:fr}) with the eq.(\ref{e:tr}) one gets
\begin{eqnarray}\label{e:last}
ma = mgsin\beta + \frac{Ia}{r^2} - f_{r} \nonumber \\
ma - \frac{Ia}{r^2}  = mg sin\beta - f_{r} \nonumber\\
 (m - \frac{I}{r^2} ) a = mgsin\beta - f_{r}
\end{eqnarray}
For the ball that we used, the value of $I$ lies between that of a
hollow sphere with a thin shell and that of a solid sphere.
Therefore the term in the brackets in the eq.(\ref{e:last}) can
never be zero. Hence, eq.(\ref{e:last}) determines the
acceleration of the ball with respect to the ground when it's
rolling down or, rolling up or, remaining stationary as the
conveyor belt beneath moves up. In the particular case, when it's
remaining in one position with respect to the ground, acceleration
is zero, leading to
\begin{equation}
mgsin\beta- f_{r}  = 0
\end{equation}
or,
\begin{equation}\label{e:fr1}
 f_{r} = mg sin \beta.
\end{equation}
Following usual convention,
\begin{equation}\label{e:n1}
f_{r} = \mu_r N
\end{equation}
where, $N$ is the normal reaction of the plane on the ball. From
the eqns.(\ref{e:n}) and (\ref{e:n1}), we get
\begin{equation}\label{e:fr2}
 f_{r} = \mu_{r} mg cos\beta
\end{equation}
From the eqns.(\ref{e:fr1}) and (\ref{e:fr2}), we get
\begin{equation}\label{e:final}
\mu_{r}   =  tan\beta.
\end{equation}
This equation(\ref{e:final}) is the basis of our experiment.
\end{section}
\begin{section}{Theoretical Prelude}
Moreover,
\begin{equation}\label{e:mu}
\mu_{r}= k_{rol} \omega_{ball}.
\end{equation}
This expression(\ref{e:mu}) relating the coefficient of rolling
friction and the angular speed of a viscoelastic ball rolling on a
hard plane, as shown in the Fig.\ref{Figure3}, was first derived
in 1996. The result was derived basing on the celebrated Hertz
contact problem \cite{{hertz},{landau}}, taking only rolling
friction loss due to the bulk into account. It was found that this
relation is linear for low velocities compared to the velocity of
sound in the body. The linearity prevails as long as the
characteristic time of the deformation process is much larger than
the dissipative relaxation time of the material. It was also found
that
\begin{equation}\label{rol}
k_{rol} =
\frac{1}{3}\frac{(3\eta_{2}-\eta_{1})^{2}}{(3\eta_{2}+2\eta_{1})}
\frac{(3E_{2}+2E_{1})}{(3E_{2}-E_{1})^{2}}.
\end{equation}
with $k_{rol}$ carrying the dimension of time. $E_{2}$ and
$\eta_{2}$ are the bulk elasticity and viscosity, whereas,
$\frac{E_{1}}{2}$ and $\frac{\eta_{1}}{2}$ are shear moduli of
elasticity and viscosity respectively, as per the conventions in
\cite{Feyn},
\cite{Mallock}.\\
We apply eq.(\ref{rol}) to find out $k_{rol}$ for the core of the
three lawn tennis balls. Since we do not have any knowledge about
the specific type of vulcanised India-rubber, we instead compute
$k_{rol}$ for the three kinds of vulcanised India-rubber dealt in
\cite{Mallock}. India-rubber, chemically, is a polymer called
Polyterpin. Vulcanisation introduces sulphur bridges connecting
different polymer chains. According to the content of the sulfur,
India-rubber gets different hue. If it contains 2.1 percent
sulphur, it appears red. It looks hard grey, if the sulphur
content is 3.8 percent. If it contains 5.7 percent sulphur, it's
colour is soft grey \cite{Mallock}. Viscoelasticity datas about
these three  kinds of rubber are also available in \cite{Mallock}.

To do the estimation, we simplify the expression for $k_{rol}$.
Moreover, we note that for an isotropic material only two elastic
constants and only two viscosity coefficients are
independent\cite{{Feyn},{landau}}. These are shear and bulk
moduli. On the top of it, if we assume isotropy of
relaxation-time, $E_{2}$ and $\eta_{2}$ have got the similar
formal expressions,
\begin{eqnarray}\label{e:bulk}
E_{2}= \frac{Y}{3(1-2\sigma)},
\eta_{2}=\frac{\overline{\eta}}{3(1-2\sigma)}.
\end{eqnarray}
Expressions for $\frac{E_{1}}{2}$ and $\frac{\eta_{1}}{2}$ are
also exactly similar,
\begin{eqnarray}\label{e:shear}
\frac{E_{1}}{2}= \frac{Y}{2(1+\sigma)}, \frac{\eta_{1}}{2}
=\frac{\overline{\eta}}{2(1+\sigma)}
\end{eqnarray}
It happens so that the relatively less discussed of all these
moduli is the Trouton viscosity\cite{{rheo}, {trou}},
$\overline{\eta}$, with the following definition,
\begin{equation}
\overline{\eta}= \frac{\frac{F}{A}}{\frac{dv_{x}}{dx}}
\end{equation}
for a viscoelastic rod with area A, applied force F, longitudinal
velocity, $v_{x}$, at a position $x$. We recall for the same rod,
Young's modulus is as
\begin{equation}
Y= \frac{\frac{F}{A}}{\frac{du_{x}}{dx}},
\end{equation}
where,
\begin{equation}
v_{x}=\frac{du_{x}}{dt}.
\end{equation}
On using the eqns.(\ref{e:bulk},\ref{e:shear}), the
expression(\ref{rol}) reduces to
\begin{equation}
k_{rol} = \frac{\overline{\eta}}{Y}.
\end{equation}
As the bulk viscosity of the vulcanised rubber \cite{Mallock}, is
very high, Young's modulus, $Y$, is almost three times that of
shear elasticity, $\frac{E_{1}}{2}$. Hence, the Trouton viscosity,
$\overline{\eta}$, is approximately three times\cite{rheo} that of
shear viscosity $\frac{\eta_{1}}{2}$. In other words, Trouton
ratio is three\cite{trou}. As a result, the final expression of
$k_{rol}$ for vulcanised rubber is
\begin{equation}\label{rubber}
k_{rol} = \frac{3\eta_{1}}{2Y}.
\end{equation}
Noting down the experimentally determined values of
$\frac{\eta_{1}}{2}$ and $Y$, from the reference\cite{Mallock},
and using the relation(\ref{rubber}), $\mathbf{k_{rol}}$ for the
viscoelastic spheres made of the three kind of rubbers, are found
as
\begin{equation}\label{e:theo}
\mathbf{k_{rol}^{red}}= \mathbf{0.036s},
\mathbf{k_{rol}^{softgrey}}= \mathbf{0.212s},
 k_{rol}^{hard grey}=0.046s
\end{equation}
\end{section}
\begin{section}{Experiment and Observations}
The measurement of velocity dependent coefficient of rolling
friction requires the measurement of coefficient of rolling
friction at different rotational speeds. The body under
observation should undergo pure rotation for measuring the
coefficients at different rotational speeds.

When the conveyor belt is moving uphill the ball would have either
a tendency to move uphill along with the belt, move downhill in a
direction opposite to the motion of the belt or, oscillate between
the two extremes (somewhere in the middle may be). In the first
case, the ball moves uphill because the rolling frictional force
is more than that of gravitation along the incline. Hence we
increase the inclination or, reduce the speed. In the second case,
the ball moves down as the gravitational force along the incline
is more than the rolling frictional force. So we reduce the
inclination or, increase the speed. In the third case, the two
opposing forces balance each other and hence the ball undergoes an
oscillatory motion initially. The ball, due to linear and
rotational air drag\cite{thom}, soon settles down at a fixed point
w.r.t the ground, making rotation at constant angular velocity and
we get the desired reading. Due to rolling friction's dissipative
role, albeit little, the ball looses angular velocity and hence
looses the rolling frictional force keeping it balanced against
gravity and after a little while goes down along the conveyor
belt.

The conveyor belt is set at different angles of inclination. For
each angle, the motor is started at the minimum possible speed and
the ball is placed on the belt. Depending on its direction of
motion, the speed of the belt is regulated so as to achieve zero
translational velocity of the ball with respect to the ground. At
such a condition, the angle of inclination of the belt, $\beta$,
and the apparent angular velocity of the ball, $\omega$, is
measured in the following way. The rotational speed of the rollers
was measured using a tachometer. The apparent angular velocity of
the ball is then calculated by dividing the value of the linear
velocity of the belt by the value of the radius of the tennis
ball. Moreover, the linear velocity of the belt is calculated by
multiplying the radius of a roller with its rpm. Hence,
\begin{equation}
\omega=\frac{\omega_{conveyor}r_{roller}}{r_{ball}}
\end{equation}
The ball is "quasirolling" \cite{sharma} as little skidding
unavoidably comes with pure rolling. The speed of the conveyor
belt is slightly higher by $v_{skid}$ relative to the speed of the
instantaneous point of contact on the ball. As a result,
\begin{equation}\label{e:skid}
\omega_{ball}=\omega - \frac{v_{skid}}{r_{ball}}.
\end{equation}
The measurement of the angle of the incline is done in a very
simplified way. The whole setup can be assumed to be a right
angled triangle with its hypotenuse representing a straight line
between the two rollers on which the conveyor belt runs. The
length of the hypotenuse was calculated by measuring the distance
between the centers of the two pulleys, using a meter scale and a
Vernier Callipers. The height of the right angled triangle can be
easily estimated by measuring the heights of the centers of both
the rollers and then taking their difference.

The above procedure is followed for a number of readings. A graph
is then plotted using Matlab between $\mu_{r}$, and the apparent
rotational speed of the ball, $\omega$, utilising
eq.(\ref{e:final}). The same steps are repeated for the other two
balls.
\begin{figure}
\includegraphics[width=21cm,height=7.8cm]{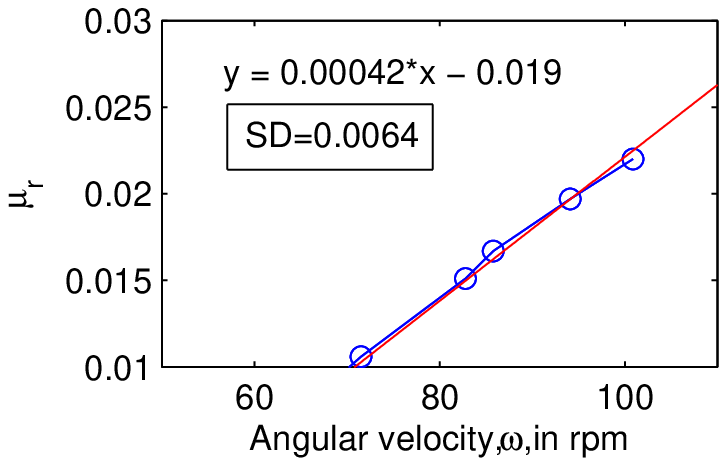}
\includegraphics[width=20cm,height=8.1cm]{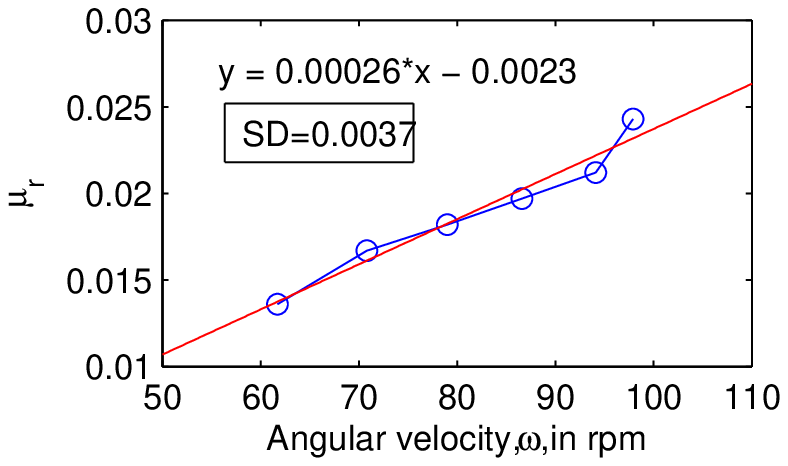}
\includegraphics[width=24.5cm,height=7.5cm]{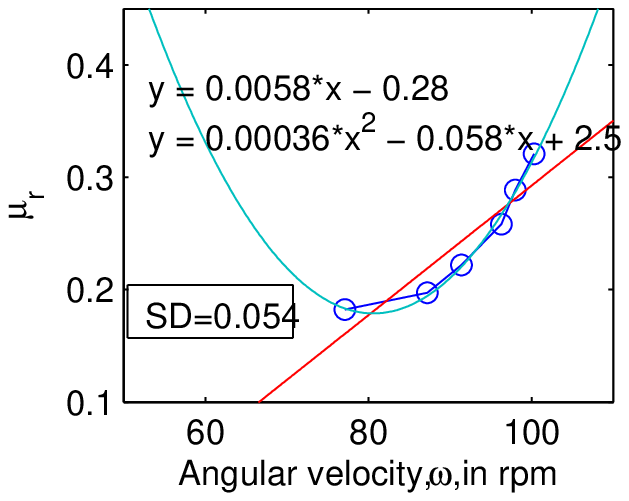}
\caption{ Graphs for a moderately old ball, \(R^2=1\); an old
ball, \(R^2 = 0.964\); a new lawn tennis ball,
\(R^2=0.821\)}\label{Figure4}
\end{figure}
It's seen that graphs are almost linear, with negative intercepts,
as expected from the eqns.(\ref{e:final}, \ref{e:mu},
\ref{e:skid}). In case of the new ball, due to the predominant
presence of felt, slipping, $v_{skid}$, quite likely have been
different for different speeds of the conveyor belt, giving rise
to the apparent quadratic nature of the plot. However, a linear
relationship is obtained in each case through least square
regression. The regression coefficient $R^2$ noted in the captions
of the graphs is the square of correlation coefficient between the
$\mu_{r}$ values observed and values on the best-fit,
\begin{equation}
R^2=1-\frac{\sum{(y_{i}-f_{i})^2}}{\sum{(y_{i}-\overline{y})^2}},
\end{equation}
where, $f_{i}$ are values on the best-fit straight line, at the
i-th position, $i=1,..,5 or,6$. Value of $R^2$ shows that for the
new ball even linear fit is good.
\begin{table}[h]
\begin{tabular}{|l|l|l|}
\hline
tennis balls &relation between $\mu_{r}$ and   & Hence,\\
             &$\omega$(in rpm)   & $\mathbf{k_{rol}}$\\
\hline
 (ages) & $\mu_r = k_{rol}\omega - const$ & (in min) \\
\hline
moderately & $\mu_r = 0.00042\omega - 0.0190$ &  0.00042  \\
old & & \\
 \hline
old & $\mu_r = 0.00026\omega - 0.0023$& 0.00026 \\
\hline
new  & $\mu_r = 0.00580\omega - 0.2800$& 0.00580 \\
\hline
\end{tabular}
\caption{Facts obtained from graphs, on experimental results, for
balls of different ages.}\label{tab:1}
\end{table}
Table \ref{tab:1} enlists the linear relations, thereof, case by
case. Hence, $k_{rol}$, determined from our rolling friction
experiment, but expressed in seconds, are as
\begin{equation}\label{e:exp}
 \mathbf{k_{rol}^{moderately old}}=
\mathbf{0.025s}, \mathbf{k_{rol}^{old}}=\mathbf{0.016s},
k_{rol}^{new}= 0.348s.
\end{equation}
\end{section}
\begin{section}{Discussion}
Comparing $R^2$ for the best-fit lines in the three graphs, we
notice that linear fit is most accurate in the case of the
moderately old ball. We see that $k_{rol}$ calculated for red-grey
rubber is also closest to $k_{rol}$ inferred from the second
graph. Moreover, on matching the relations(\ref{e:theo}) and
(\ref{e:exp}), we observe that $k_{rol}$ for the moderately old
ball/ old ball is too close to that of red grey India-rubber,
whereas $k_{rol}$ for the new ball is close to the soft-grey
India-rubber. These lead us to conclude the following three
possibilities, a) moderately old ball/old ball, we have used had
the core made up of red grey India-rubber and the new ball had the
core made up of soft-grey India-rubber; b) all the balls were
having the core made up of soft-grey India-rubber. With mechanical
degradation associated with shots, $\overline{\eta}$ and $Y$ got
reduced thus reducing $k_{rol}$ and the moderately old ball/old
ball behaving like having core made up of red-grey India-rubber;
c)all the balls used by us were having core made up of red-grey
India-rubber and for the new ball, predominant contribution is due
to felt.\\
We have used balls, available in local market. Most likely, those
were trainer balls made not following ITF specifications fully.
Moreover, ITF ball specification points to \cite{itf} vulcanised
red-grey India-rubber as the core material. In future, we plan to
take a full investigation with ITF approved professional balls,
along the line of this paper.\\
In principle, rolling friction due to surface interaction should
be taken into account. One can include rotational air drag also.
The side of the ball coming down to onto the conveyor belt has
lower pressure head in the adjoining air volume, compared to the
other side. This leads to a drag force, working opposite to
$f_{r}$ shown in the Fig.\ref{Figure3}. In that case we would have
gotten the same eq.(\ref{e:final}) with R.H.S. of the
eq.(\ref{e:mu}) getting modified by the additional presence of
rotational terms and surface rolling friction terms.\\
One limitation of this setup in this present form that it requires
at least five persons to do the experiment. The most time
consuming step of the experiment was the angle adjustment step of
the conveyor belt. We have taken each reading twice, due to
limited availability of tachometer. We are pretty sure about the
repeatability.\\
We have not used anything to reduce air hindrance, by putting a
plastic frame, say as in Ko et.al's\cite{ko} experiment.
Fabrication of our setup has cost us approximately $250$ USD. The
motor being a 150 rpm motor, we could achieve only limited ball
angular velocity. It would be interesting to get DC motors of
different hps, custom-made, and achieve higher speeds for the
conveyor belt and thereby get rolling friction coefficient, at
non-linear angular
velocity regime\cite{{xu},{xu1},{tennisdesign}}, checked for different objects.\\
However, the relation,
\begin{equation}
 \overline{\eta}=k_{rol}Y,
\end{equation}
leads us to an almost direct way of measuring Trouton viscosity of
a viscoelastic substance. Only thing we need to know as an input
from outside, the Young modulus, $Y$. Hence, the setup which we
are using to measure $k_{rol}$ directly, can be used to determine,
in addition, Trouton viscosity of a viscoelastic material.
\end{section}
\begin{section}{Acknowledgement}
We would like to thank G. J. Desai, V. V. Chaudhari and P. R.
Dhimaan for helping us design the set up, Harshit Khurana for his
comments on the manuscript and the Electrical Department Lab
personnel of BITS-GOA, for lending us a tachometer. We are
grateful to the anonymous Referees for their invaluable
suggestions.
\end{section}

\end{document}